\documentclass[final,5p,times,twocolumntwocolumns]{elsarticle}

\usepackage{amssymb}
\usepackage{graphicx}
\usepackage{dcolumn}
\usepackage{bm}
\usepackage{booktabs}
\usepackage{adjustbox}
\usepackage{blindtext}
\usepackage{amsmath}
\usepackage{color}
\usepackage{geometry}
\begin{document}

\begin{frontmatter}

\title{Robustness of chiral symmetry in atomic nuclei with reflection-asymmetric shapes}
\author{C. M. Petrache}
\address{Universit\'e Paris-Saclay, CNRS/IN2P3, IJCLab, 91405, Orsay, France}
\begin{abstract}
The present paper is a comment regarding the robustness of the chirality in presence of the space-reflection asymmetry, which leads to pairs of interleaved positive- and negative-parity bands. The recent results reported in Ref. \cite{2006.12062}  which introduced the $chiture$ and $chiplex$ quantum numbers to describe an ideal nuclear system with simultaneous chiral and reflection symmetry breaking, are commented.  
\end{abstract}
 \date{}
%----------------------------------------------------------------------------------------------------
\end{frontmatter}

Nearly degenerate quantum states arise from fundamental symmetries, which are often broken in finite many-body systems like atomic nuclei. Chirality can be static, like in molecules composed of more than four atoms, or dynamic, like in particle physics distinguishing between parallel and antiparallel orientation of the spin with respect to the momentum of massless fermions. Frauendorf and Meng \cite{Stefan-Jie} pointed out that the rotation of an axial-asymmetric (triaxial) nucleus can attain a chiral character: the angular momentum vector $\vec J$ introduces chirality by selecting different left-handed or right-handed systems in the intrinsic frame of reference of a triaxial nucleus, giving rise to degenerate rotational bands because all octants are energetically equivalent. The nuclear chirality is therefore dynamic, resulting from the combination of dynamics induced by the angular momentum, and geometry induced by the triaxial shape. More than 60 chiral doublet bands have been reported in $A \sim$ 80, 100, 130, and 190 mass regions \cite{Xiong2019}, which are regarded as fingerprints of triaxial deformation. 

An important advance in the understanding of the chiral bands in triaxial nuclei with octupole correlations was reported in Ref. \cite{2006.12062}. The spontaneous breaking of the space-reflection symmetry gives rise to a pair of interleaved positive- and negative-parity bands \cite{butler1996}, while the breaking of the chiral symmetry gives rise to a pair of nearly degenerate rotational bands for each of the positive- and negative-parity bands, leading thus to a set of four nearly degenerate bands. The $chiture$ and $chiplex$ quantum numbers are introduced for describing an ideal nuclear system with simultaneous chiral and reflection symmetry breaking. The new symmetries are used to categorize the excited energy states and to establish the selection rules of the corresponding electromagnetic transitions. These findings can be considered as critical signals for the presence of chirality in pear-shaped nuclei and trigger future experimental investigations.

The shapes of real nuclei can be described by a superposition of multipoles corresponding to triaxial and octupole shapes, associated to chiral and space-reflection symmetries, leading to chiral and parity doublets, respectively. In addition, certain deformed nuclei can have nearly degenerate bands with the same parity, which can be associated to the pseudo-spin symmetry \cite{Liang2015}. Reflection asymmetry, static or dynamic, is observed in nuclei when  $\Delta l=3$ orbitals are close to the Fermi surface, like for example the proton $h_{11/2}$ and $d_{5/2}$ orbitals in the $A\approx130$ region, or the neutron $i_{13/2}$ and $f_{7/2}$ orbitals in the $A\approx150$ region. In such cases the octupole degree of freedom may play an important role and the nuclei can acquire a pear shape. Pseudo-spin symmetry is observed in nuclei  when nearly degenerate $\Delta l=2$ orbitals are close to the Fermi surface, like for example the proton $d_{5/2}$ and $g_{7/2}$ orbitals in the $A\approx130$ region, or the neutron $s_{1/2}$ and $d_{3/2}$ orbitals in the $A\approx150$ region. It is therefore possible to observe quartet bands, that is four nearly degenerate bands, having positive and negative parities in nuclei with reflection-asymmetric shapes, and quartet bands having the same parity in deformed nuclei with active pseudo-spin orbitals. The nuclei with $A \approx 130$ are of special interest because the proton Fermi surface is close to both   $\Delta l=3$ proton orbitals ($h_{11/2}$, $d_{5/2}$)  leading to reflection asymmetry and $\Delta l=2$ proton orbitals ($d_{5/2}$ and $g_{7/2}$) leading to pseudo-spin symmetry.
The complex and intriguing situation of three broken symmetries in a single nucleus - chiral, space-reflection and pseudo-spin - can occur, giving rise to several nearly degenerate bands. Multiple chiral doublet bands have been recently predicted (M$\chi$D) \cite{Meng2006}, triggering several experimental studies which led to the observation of several nearly degenerate bands built on two or more quasiparticle configurations \cite{133Ce, 103Rh, 78Br,136Nd, 135Nd,131Ba}. However, from the theoretical point of view, the description of the chiral bands in triaxial nuclei with reflection-asymmetric shapes involving in addition pseudo-spin orbitals,  is not an easy task. A first step was recently achieved by developing the reflection-asymmetric triaxial particle rotor model (RAT-PRM) \cite{WANG2019454}. The model was applied for the first time to an ideal case of an odd-odd nucleus with maximum triaxial deformation $\gamma=90^{\circ}$, static octupole deformation with energy of the lowest negative-parity state $E(0^-)=0$ MeV, a pure configuration with two-$j$ shells ($h_{11/2}$ and $d_{5/2}$) having the orbital and total angular momenta differing by $3 \hbar$ ($\Delta l=\Delta j=3\hbar$), and energy spacing typical for $A \approx 130$ nuclei \cite{2006.12062}. The energy spectra, electromagnetic transition probabilities and chiral geometry of the nearly degenerate quartet bands have been investigated. A new symmetry was derived for maximum triaxiality ($\gamma=90^{\circ}$), characterized by a new quantum number $\cal A$ called $chiture$ in analogy with the $signature$ quantum number $r=e^{-i\alpha\pi}$ of the operator $\mathcal{R}_z(\pi)$ describing the rotation by an angle $\pi$ about the $z$-axis \cite{Frau-RMP}, or the new quantum number $\cal B=\cal A \cal P$ called $chiplex$, in analogy with the $simplex$ quantum number $\cal S=\cal R(\pi)\cal P$, which is obtained from the product of the chiture $\cal A$ and parity $\cal P$ quantum numbers. The states with same parity $\cal P$ but different chiture $\cal A$ constitute the chiral doublets, while the states with same chiture $\cal A$ but different parity $\cal P$ constitute the parity doublets.The selection rules for in-band and out-of-band $E2, M1, E3$ transition probabilities have been revealed, showing that they only occur  between states with different chiplex $\cal B$. The $M1$ transitions exhibit a staggering behavior as function of spin, while the $E3$ transitions alternate with spin. The static chirality regime is investigated using the {\it azimuthal plot}.

The work reported in Ref. \cite{2006.12062} represents a significant progress in the immense effort to built a model which can describe the complex band structure resulting from the Chirality-Parity (ChP) violation in triaxial nuclei with reflection asymmetry. These theoretical investigations emphasize the robustness of the chiral motion in nuclei, which is not destroyed by octupole correlations. Future extensions of RAT-PRM for the description of three- and four-quasiparticle configurations in odd-even and even-even nuclei, respectively, as well as the inclusion of multi-$j$ shells for the description of bands built on pseudo-spin partners, are next steps that the Peking theory group can afford with success.

% \begin{figure*}[!htbp]
% \hskip -. cm
%\vskip -. cm
%\rotatebox{90}{\scalebox{1.0}{\includegraphics{levelscheme_new.eps}}}
%\vskip -. cm
%  \begin{adjustbox}{addcode={
%    \begin{minipage}{\width}}{
%\caption {\label{fig1} (Color online) Partial level scheme of  $^{131}$Ba deduced from the present work. Transition energies are given in keV and their measured relative intensities are proportional to the widths of the arrows. The labels of the newly identified transitions are on white background, while those of the previously known transitions are on a clolored background. Levels and intraband transitions are colored in group by bands. The energies and ends of linking transitions are colored by their initial states while the tips are colored by their final states. Uncertain spin and parity assignments are given in brackets.}
%\end{minipage}},rotate=90,center}
%\includegraphics[scale=1.1]{levelscheme_new.eps}
%\end{adjustbox}
%\end{figure*}

\end{document}